\documentclass[twocolumn,nofootinbib,amsmath,amssymb,a4paper]{revtex4}

\usepackage{graphicx}
\usepackage{dcolumn}
\usepackage{bm}

\newcommand{\gsim}{
\mathrel{\hbox{\rlap{\hbox{\lower4pt\hbox{$\sim$}}}\hbox{$>$}}}}

\begin{document}

\begin{flushright}
\vspace*{-0.9truecm}
CERN-PH-TH/2007-016
\end{flushright}

\vspace*{-1.7truecm}

\title{\boldmath Theoretical Review of $\gamma/\phi_3$ Measurements with
$B_s$ Decays to Charm\unboldmath}

\author{Robert Fleischer}
 \email{Robert.Fleischer@cern.ch}
\affiliation{%
Theory Division, Physics Department, CERN, CH-1211 Geneva 23,
Switzerland
}%

\begin{abstract}
We give an overview of various determinations of $\gamma/\phi_3$ with the help 
of $B_s$ decays into charmed final states, distinguishing between transitions 
with tree and penguin contributions and pure tree decays. In the corresponding
strategies, the $U$-spin flavour symmetry of strong interactions provides a very 
useful tool, and offers interesting ``by-producs" for the $B$-physics programme
at the LHC, including the control of the penguin uncertainties in the determinations
of the $B^0_d$--$\bar B^0_d$ and $B^0_s$--$\bar B^0_s$ mixing phases 
$\phi_d$ and $\phi_s$ from $B_d\to J/\psi K_{\rm S}$ and $B_s\to J/\psi\phi$, 
respectively, and an alternative extraction of the latter phase through 
$B_s\to D_s^+D_s^-$. Finally, we point out that the cleanest determinations of 
the mixing phases $\phi_s$ and $\phi_d$ are offered by the pure tree decays
$B_s\to D_\pm K_{\rm S(L)}$ and $B_d\to D_\pm\pi^0, D_\pm\rho^0, ...$, respectively,
which are very interesting for the searches of new physics.
\end{abstract}

\maketitle

\section{Setting the Stage}\label{sec:intro}
During the recent years, we have seen a tremendous progress in $B$ physics  
\cite{ferroni}. The data agree globally with the Kobayashi--Maskawa (KM) mechanism 
of CP violation, i.e.\ with the Standard Model (SM).  However, we have also hints for 
discrepancies, which could be first signals of new physics (NP). Unfortunately, 
the uncertainties are still too large to draw firm conclusions. 

Thanks to the start of the LHC, exciting new perspectives will arise in the autumn of 
2007, also for $B$-decay studies \cite{nakada}. The LHCb experiment will allow 
us to fully exploit the physics potential of the $B_s$-meson system, and precision determinations of the angle $\gamma$ of the unitarity triangle will become possible,\footnote{In the following, I shall use the $\gamma(\equiv\phi_3)$,
$\beta(\equiv\phi_1)$ notation.}
which are a key ingredient for the search of NP in the flavour sector. 

Let us therefore have a closer look at $B_s$ decays into final states with charm, 
which is the topic of WG5 of the CKM2006 workshop, where this talk was given. 
The notation and formulae for the CP asymmetries in decays of neutral $B_q$ 
mesons ($q\in\{d,s\}$) are given as follows \cite{RF-rev}:
\begin{eqnarray}
\lefteqn{\frac{\Gamma(B^0_q(t)\to f)-
\Gamma(\bar B^0_q(t)\to \bar f)}{\Gamma(B^0_q(t)\to f)+
\Gamma(\bar B^0_q(t)\to \bar f)}}\nonumber\\
&&=\left[\frac{{\cal A}_{\rm CP}^{\rm dir}\,\cos(\Delta M_q t)+{\cal A}_{\rm CP}^{\rm mix}\,
\sin(\Delta M_q t)}{\cosh(\Delta\Gamma_qt/2)-{\cal A}_{\rm \Delta\Gamma}\,
\sinh(\Delta\Gamma_qt/2)}\right],\label{ACP-timedep}
\end{eqnarray}
where 
\begin{equation}
{\cal A}_{\rm CP}^{\rm dir}=\frac{1-|\xi_f^{(q)}|^2}{1+|\xi_f^{(q)}|^2} 
\quad\mbox{and}\quad 
{\cal A}_{\rm CP}^{\rm mix}=\frac{2\,\mbox{Im}\,{\xi^{(q)}_f}}{1+|\xi^{(q)}_f|^2}
\end{equation}
denote the ``direct" and ``mixing-induced" CP-violating observables, respectively.
The quantity ${\cal A}_{\rm \Delta\Gamma}$, which should be accessible in the 
$B_s$-meson system because of the expected sizeable width difference 
$\Delta\Gamma_s$, satisfies 
\begin{equation}
({\cal A}_{\Delta\Gamma})^2=1-
({\cal  A}^{\mbox{{\scriptsize dir}}}_{\mbox{{\scriptsize CP}}})^2-
({\cal  A}^{\mbox{{\scriptsize mix}}}_{\mbox{{\scriptsize CP}}})^2.
\end{equation}
These observables are governed by 
\begin{equation}
\xi_f^{(q)} \sim
e^{-i \phi_q}
\left[\frac{A(\overline{B_q^0}\to f)}{A(B_q^0\to f)}\right],
\end{equation}
where the $B^0_q$--$\bar B^0_q$ mixing phases
\begin{equation}
\phi_q\stackrel{\rm SM}{=}2\,\mbox{arg}(V_{tq}^\ast V_{tb})=
\left\{
\begin{array}{cc}
+2\beta & (q=d)\\
-2\lambda^2\eta & (q=s)
\end{array}
\right.
\end{equation}
are an important input for the following discussion. The data for the CP violation in
$B^0_d\to J/\psi K_{\rm S}$ (and similar decays) imply 
$\phi_d=(42.4\pm2)^\circ$ \cite{HFAG}. Performing a measurement of the
untagged, time-dependent three-angle distribution of the 
$B^0_s\to J/\psi[\to \ell^+\ell^-]\phi[\to K^+K^-]$ decay products
\cite{DDFN}, the D0 collaboration
has recently reported the following result \cite{D0}:
\begin{equation}
\phi_s=-0.79\pm0.56\,\mbox{(stat.)}\pm0.01\,\mbox{(syst.).}
\end{equation}
Consequently, this phase is still not stringently constrained. However, it is 
very accessible at the LHC \cite{nakada}. The determinations of the 
$\phi_q$ work also in the presence of CP-violating NP contributions to 
$B^0_q$--$\bar B^0_q$ mixing, provided we have negligible contributions of this 
kind to the corresponding decay amplitudes, which is a very plausible (and testable, 
see below) assumption for the $B^0_d\to J/\psi K_{\rm S}$ and $B^0_s\to J/\psi\phi$ 
decays. For the following discussion, we hence assume that the  $\phi_q$ are known.

\section{Decays with Tree and Penguin Contributions}\label{sec:tree-pen}
In this section, we have a fresh look at the strategies proposed in
\cite{RF-BspsiK-DD}. Here the $U$-spin flavour symmetry of strong interactions, 
which relates down and strange quarks in the same way as the isospin symmetry 
relates down and up quarks, is used to extract $\gamma$ from 
$B_{d,s}$ decays with tree and penguin contributions. The conceptual advantage 
of these $U$-spin strategies is -- in contrast to the ``conventional" $SU(3)$ 
flavour-symmetry strategies -- that no additional dynamical assumptions, such
as the neglect of annihilation topologies, have to be made. 

\boldmath
\subsection{The $B_{s(d)}\to J/\psi K_{\rm S}$ System}
\unboldmath
In the SM, the decay amplitudes of $B^0_s\to J/\psi K_{\rm S}$
and $B^0_d\to J/\psi K_{\rm S}$ channels can be written as follows:
\begin{eqnarray}
A(B^0_s\to J/\psi K_{\rm S})&=&-\lambda{\cal A}\left[1-ae^{i\theta}e^{i\gamma}
\right]\\
A(B^0_d\to J/\psi K_{\rm S})&=&
\left(1-\frac{\lambda^2}{2}\right){\cal A}'\left[1+\epsilon a'e^{i\theta'}e^{i\gamma}
\right]
\end{eqnarray}
where $\epsilon\equiv\lambda^2/(1-\lambda^2)=0.05$, and 
\begin{equation}\label{label-Hadr}
{\cal A}=\mbox{``tree+pen"},\quad ae^{i\theta}=\mbox{``pen/(tree+pen)''}
\end{equation}
are CP-conserving strong quantities. The untagged rates
\begin{equation}
\langle \Gamma(B\to f) \rangle\equiv 
\Gamma(B(t)\to f)+\Gamma(\bar B(t)\to f)
\end{equation}
allow us to introduce
\begin{eqnarray}
H&\equiv&\mbox{PhSp}\times\frac{1}{\epsilon}
\left|\frac{{\cal A}'}{{\cal A}}\right|^2
 \frac{\langle \Gamma(B_s\to J/\psi K_{\rm S})
\rangle}{\langle \Gamma(B_d\to J/\psi K_{\rm S}) \rangle}\nonumber\\
&=&\frac{1-2a\cos\theta\cos\gamma+a^2}{1+2\epsilon
a'\cos\theta'\cos\gamma+ \epsilon^2 a'^2},\label{H-def}
\end{eqnarray}
where PhSp denotes a straightforward phase-space factor. On the other hand, 
tagged, time-dependent rate measurements allow us to extract
\begin{eqnarray}
{\cal A}_{\rm CP}^{\rm dir}(B_s\to J/\psi K_{\rm S})&=&F_1(a,\theta,\gamma)\\
{\cal A}_{\rm CP}^{\rm mix}(B_s\to J/\psi K_{\rm S})&=&
F_2(a,\theta,\gamma,\phi_s),
\end{eqnarray}
as well as
\begin{equation}
{\cal A}_{\rm CP}^{\rm dir}(B_d\to J/\psi K_{\rm S})=
F'_1(\epsilon a',\theta',\gamma)=0+{\cal O}(\epsilon a')
\end{equation}
\vspace*{-0.6truecm}
\begin{eqnarray}
{\cal A}_{\rm CP}^{\rm mix}(B_d\to J/\psi K_{\rm S})&=&
F'_2(\epsilon a',\theta',\gamma,\phi_d)\nonumber\\
&=&-\sin\phi_d+{\cal O}(\epsilon a').\label{ACPmix-BdpsiK}
\end{eqnarray}
It is an important feature of the $B_{s(d)}\to J/\psi K_{\rm S}$
system that the corresponding decays are related through the interchange
of all down and strange quarks. The $U$-spin flavour symmetry of strong interactions 
hence implies:
\begin{equation}\label{A-rel}
|{\cal A}'|=|{\cal A}|;
\end{equation}
\vspace*{-0.8truecm}
\begin{equation}\label{a-thet-rel}
a'=a, \quad \theta'=\theta.
\end{equation}
Consequently, (\ref{A-rel}) and (\ref{a-thet-rel}) allow us to determine the
quantity $H$ introduced in (\ref{H-def}).  Here $U$-spin-breaking corrections to 
(\ref{A-rel}) have the most important impact:
\begin{equation}
\left|\frac{{\cal A}'}{{\cal A}}\right|_{\rm fact}=
\frac{F_{B^0_dK^0}(M_{J/\psi}^2;1^-)}{F_{B^0_d\bar K^0}(M_{J/\psi}^2;1^-)},
\end{equation}
whereas corrections to (\ref{a-thet-rel}) play a minor r\^ole because of the
$\epsilon$ suppression in (\ref{H-def}). Finally, $\gamma$, $a$, $\theta$ can 
be extracted from $H$, ${\cal A}_{\rm CP}^{\rm dir}(B_s\to J/\psi K_{\rm S})$ and
${\cal A}_{\rm CP}^{\rm mix}(B_s\to J/\psi K_{\rm S})$.

\begin{figure}
\includegraphics[width=0.27\textwidth]{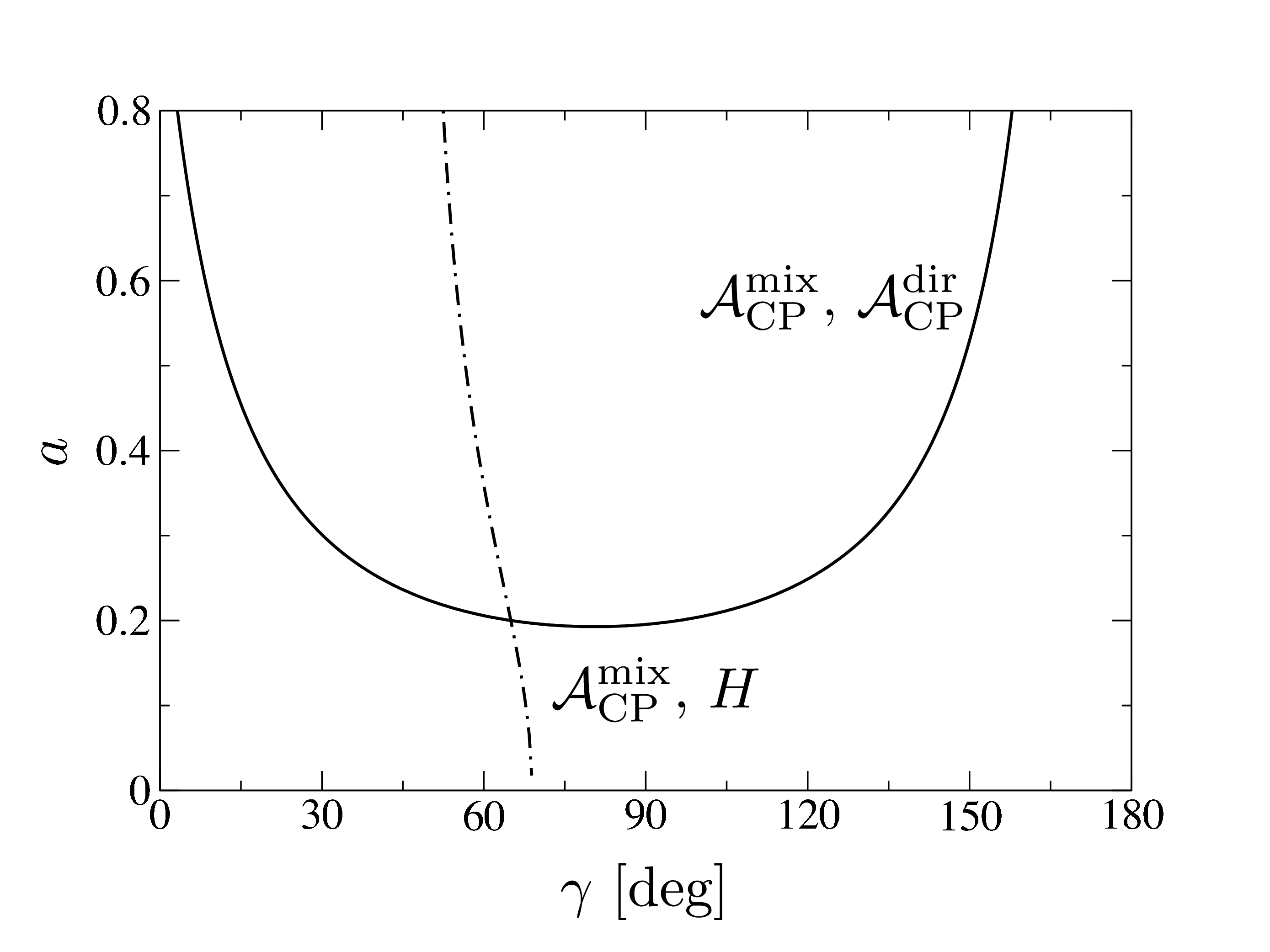}
\vspace*{-0.2truecm}
\caption{\label{fig:BspsiK-cont} Illustration of the extraction of $\gamma$ and $a$
from the $B_{s(d)}\to J/\psi K_{\rm S}$ system through contours in the $\gamma$--$a$
plane.}
\end{figure}

In Fig.~\ref{fig:BspsiK-cont}, we illustrate this determination: using $\gamma=65^\circ$, 
$\phi_s=-2^\circ$, $a=a'=0.2$ and $\theta=\theta'=30^\circ$ as input parameters, 
we obtain ${\cal A}_{\rm CP}^{\rm dir}(B_s\to J/\psi K_{\rm S})=0.20$,
${\cal A}_{\rm CP}^{\rm mix}(B_s\to J/\psi K_{\rm S})=0.35$ and $H=0.89$. These
observables can be converted into the contours shown in the figure, where the one
following from the CP asymmetries ${\cal A}_{\rm CP}^{\rm dir}$ and 
${\cal A}_{\rm CP}^{\rm mix}$ is {\it theoretically} clean. The intersection of the 
two contours allows a transparent extraction of $\gamma$ and $a$ (we 
have excluded additional solutions arising for unphysically large values of $a>1$). 

Another interesting aspect of this strategy is that we have so far not used the
$B^0_d$--$\bar B^0_d$ mixing phase $\phi_d$. The $U$-spin relation (\ref{a-thet-rel})
allows us -- in combination with the extracted values of $\gamma$, $a$ and $\theta$ --
to control the penguin uncertainties affecting the determination of $\phi_d$
from (\ref{ACPmix-BdpsiK}). These corrections received quite some attention
\cite{MishCiu}, and can actually be controlled at the LHC through the data for 
the $B_s\to J/\psi K_{\rm S}$ channel. For the practical implementation, also the
$U$-spin relation
\begin{equation}
{\cal A}_{\rm CP}^{\rm dir}(B_d\to J/\psi K_{\rm S})=- \epsilon H
{\cal A}_{\rm CP}^{\rm dir}(B_s\to J/\psi K_{\rm S})
\end{equation}
is very useful.

The strategy discussed above has also a counterpart in the $B_s$-meson system. 
As we noted above, $\phi_s$ can be extracted from $B_s\to J/\psi \phi$ \cite{DDFN}. 
Since we have two vector mesons in the final state, the CP eigenstates have 
to be disentangled through the $J/\psi[\to \ell^+\ell^-]$,
$\phi [\to K^+K^-]$ angular distribution. Here it is convenient to introduce linear 
polarization states $f\in$ $\{0$, $\parallel$ (CP even); $\perp$ (CP odd)$\}$,
which allow us to write
\begin{equation}
\xi^{(s)}_{(\psi\phi)_f}\,\propto\, e^{-i\phi_s}
\biggl[1-2\,i\,\lambda^2 a_f' e^{i\theta_f'}
\sin\gamma + {\cal O}(\lambda^4)\biggr].
\end{equation}
As $\phi_s\approx -2^\circ$ in the SM, the penguin effects have 
a significant impact in this case, at least at the 20\% level (which may be
enhanced through final-state interaction effects). Therefore, the question of 
how to control these effects arises, which is particularly relevant for
LHCb upgrade plans. These uncertainties can actually be controlled through an
analysis of the $B_d\to J/\psi \rho^0$, $B_s\to J/\psi \phi$ system \cite{RF-ang}:
using $\phi_d$ and the $U$-spin symmetry, $\gamma$ and the hadronic
parameters $(a_f', \theta_f')$ can be extracted, allowing us to include
the penguins in the extraction of $\phi_s$.

\boldmath
\subsection{The $B_{d(s)}\to D^+_{d(s)}D^-_{d(s)}$ System}
\begin{figure}
\includegraphics[width=0.27\textwidth]{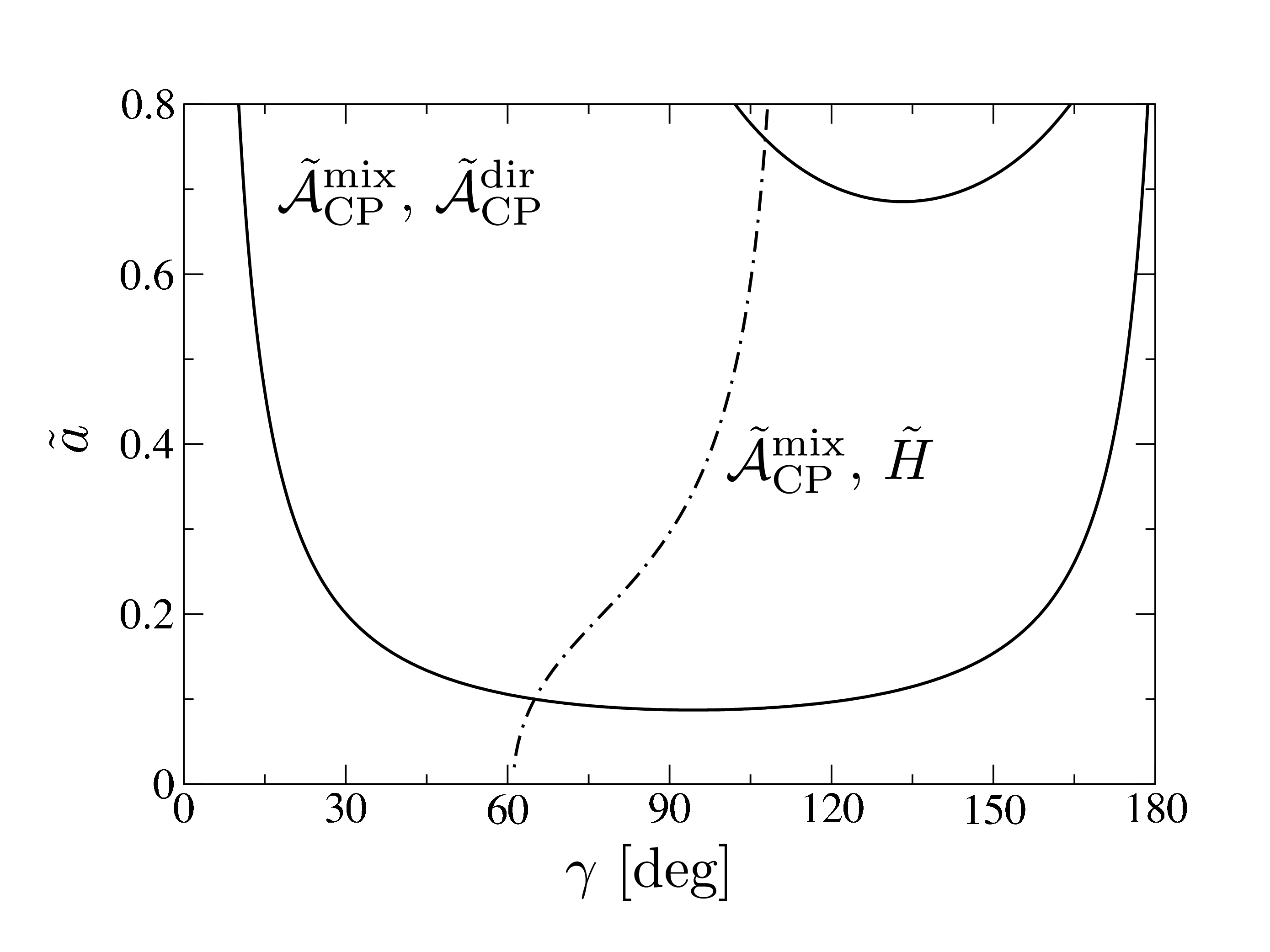}
\vspace*{-0.2truecm}
\caption{\label{fig:BDD-cont} Illustration of the extraction of $\gamma$ and $\tilde a$
from the $B_{s(d)}\to D^+_{s(d)} D^-_{s(d)}$ system
through contours in the $\gamma$--$\tilde a$ plane.}
\end{figure}

\unboldmath
In the SM, the  decay amplitudes take the form
\begin{eqnarray}
A(B^0_d\to D_d^+D_d^-)&\hspace*{-0.1truecm}=&
\hspace*{-0.1truecm}-\lambda \tilde {\cal A}\left[1-\tilde a
e^{i\tilde\theta}e^{i\gamma}\right]\\
A(B^0_s\to D_s^+D_s^-)&\hspace*{-0.25truecm}=&\hspace*{-0.3truecm}
\left(1-\frac{\lambda^2}{2}\right) \tilde {\cal A}'
\left[1+\epsilon\tilde a'e^{i\tilde\theta'}e^{i\gamma}\right]\hspace*{-0.1truecm},
\end{eqnarray}
where $\tilde {\cal A}$ and $\tilde a e^{i\tilde\theta}$ are defined in analogy to 
(\ref{label-Hadr}). We may then, as in (\ref{H-def}), introduce a quantity 
\begin{eqnarray}
\tilde H&\equiv& \mbox{PhSp}\times\frac{1}{\epsilon}
\left|\frac{\tilde {\cal A}'}{\tilde {\cal A}}\right|^2
 \frac{\langle \Gamma(B_d\to D_d^+D_d^-)
\rangle}{\langle \Gamma(B_s\to D_s^+D_s^-) \rangle}\nonumber\\
&=&\frac{1-2\tilde a\cos\tilde \theta\cos\gamma+\tilde a^2}{1+2 \epsilon 
\tilde a'\cos\tilde \theta'\cos\gamma+ \epsilon^2\tilde a'^2}.\label{Htilde-def}
\end{eqnarray}
Furthermore, we may write
\begin{eqnarray}
{\cal A}_{\rm CP}^{\rm dir}(B_d\to D_d^+D_d^-)&=&F_1(\tilde a,\tilde \theta,
\gamma)\\
{\cal A}_{\rm CP}^{\rm mix}(B_d\to D_d^+D_d^-)&=&
F_2(\tilde a,\tilde \theta,\gamma,\phi_d),
\end{eqnarray}
as well as
\begin{equation}
{\cal A}_{\rm CP}^{\rm dir}(B_s\to D_s^+D_s^-)=
F'_1(\epsilon \tilde a',\tilde \theta',\gamma)=
0+{\cal O}(\epsilon \tilde a')
\end{equation}
\vspace*{-0.2truecm}
\begin{eqnarray}
{\cal A}_{\rm CP}^{\rm mix}(B_s\to D_s^+D_s^-)&=&
F'_2(\epsilon \tilde a',\tilde \theta',\gamma,\phi_s)\nonumber\\
&=&-\sin\phi_s+{\cal O}(\epsilon\tilde a').\label{AmixBsDsDs}
\end{eqnarray}
Since the $B_{d(s)}\to D^+_{d(s)} D^-_{d(s)}$ decays are again related through 
the interchange of all $d$ and $s$ quarks, the $U$-spin flavour symmetry of strong 
interactions implies
\begin{equation}\label{A-rel-D}
|\tilde {\cal A}'|=|\tilde {\cal A}|;
\end{equation}
\vspace*{-0.3truecm}
\begin{equation}\label{a-thet-rel-D}
\tilde a'=\tilde a, \quad \tilde \theta'=\tilde \theta.
\end{equation}
Consequently, (\ref{A-rel-D}) and (\ref{a-thet-rel-D}) allow us to determine 
$\tilde H$, where the $U$-spin-breaking corrections to (\ref{A-rel-D}) have the most important impact:
\begin{equation}
\left|\frac{\tilde {\cal A}'}{\tilde {\cal A}}\right|_{\rm fact}
\hspace*{-0.5truecm}\approx
\frac{(M_{B_s}-M_{D_s})\,\sqrt{M_{B_s}M_{D_s}}\,(w_s+1)}{(M_{B_d}-M_{D_d})
\,\sqrt{M_{B_d}M_{D_d}}\,(w_d+1)}\frac{f_{D_s}\,\xi_s(w_s)}{f_{D_d}\,
\xi_d(w_d)}.
\end{equation}
Because of the $\epsilon$ suppression in (\ref{Htilde-def}), corrections to 
(\ref{a-thet-rel-D}) play a minor r\^ole. Finally, we may extract $\gamma$, $\tilde a$ 
and $\tilde \theta$ from the measured values of $\tilde H$, 
${\cal A}_{\rm CP}^{\rm dir}(B_d\to D_d^+D_d^-)$ and
${\cal A}_{\rm CP}^{\rm mix}(B_d\to D_d^+D_d^-)$, as illustrated in
Fig.~\ref{fig:BDD-cont}. In this example, we have used the input parameters
$\gamma=65^\circ$, $\phi_d=42.4^\circ$, $\tilde a=\tilde a'=0.1$ and 
$\tilde \theta=\tilde \theta'=210^\circ$, which yield the observables
${\cal A}_{\rm CP}^{\rm dir}(B_d\to D_d^+D_d^-)=-0.08$,
${\cal A}_{\rm CP}^{\rm mix}(B_d\to D_d^+D_d^-)=0.78$ and
$\tilde H=1.1$. It should be noted that the contour following from 
$\tilde {\cal A}_{\rm CP}^{\rm dir}$, $\tilde {\cal A}_{\rm CP}^{\rm mix}$ is 
{\it theoretically} clean.

An interesting by-product of this strategy is that we can now also control the 
penguin effects in the extraction of $\phi_s$ from (\ref{AmixBsDsDs}). This
avenue provides an attractive alternative to the conventional determination 
of the $B^0_s$--$\bar B^0_s$ mixing phase through the $B_s\to J/\psi \phi$
angular analysis.

\section{Pure Tree Decays}\label{sec:tree}
The determination of $\gamma$ from $B\to D^{(*)}K^{(*)}$ tree decays suffers 
currently from large uncertainties:
\begin{equation}
\left.\gamma\right|_{D^{(*)} K^{(*)}} = \left\{
\begin{array}{ll}
(62^{+38}_{-24})^\circ & \mbox{(CKMfitter)}\\[5pt] 
(82\pm 20)^\circ & \mbox{(UTfit).}
\end{array}\right.
\end{equation}
This unfavourable situation can be significantly improved at LHCb \cite{nakada}, 
where also $B_s$ decays into final states with charm play an important r\^ole. 

\boldmath
\subsection{The $B_s\to D_s^\pm K^\mp$, $B_d\to D^\pm\pi^\mp$ System}
\unboldmath
The key feature of these decays is that both a $B^0_q$ and a $ \bar B^0_q$ meson
may decay into the same final state $D_q\bar u_q$, thereby leading to interference
between mixing and decay processes. This phenomenon brings $\phi_q$ and
$\gamma$ into the game, which enter actually in the combination 
$\phi_q+\gamma$. In the case of $q=s$, corresponding to the final states
$D_s\in\{D_s^+, D_s^{\ast+}, ...\}$ and $u_s\in\{K^+, K^{\ast+}, ...\}$, the interference
effects are governed by a hadronic parameter $X_se^{i\delta_s}\propto R_b$, where
$R_b$ is the usual side of the unitarity triangle, and are hence large. On the other hand, 
for $q=d$ with $D_d\in\{D^+, D^{\ast+}, ...\}$ and $u_d\in\{\pi^+, \rho^+, ...\}$, 
a doubly Cabibbo-suppressed hadronic parameter 
$X_de^{i\delta_d}\propto -\lambda^2R_b$ enters, which leads to tiny interference 
effects.

If the $\cos(\Delta M_qt)$ and $\sin(\Delta M_qt)$ terms of the time-dependent 
decay rates are measured, $\phi_q+\gamma$ can be extracted in a theoretically
clean way \cite{clean-gam}. Since the mixing phases $\phi_q$ are known, this
determination can be converted into a measurement of $\gamma$. However, 
the practical implementation is affected by problems: we encounter an 
eightfold discrete ambiguity for $\phi_q+\gamma$, which has to be resolved for the 
search of NP. Moreover, in the $q=d$ case, an additional input is required to 
extract $X_d$, since ${\cal O}(X_d^2)$ interference effects would have to be 
resolved, which is impossible. 

As was pointed out in \cite{RF-Bdpi}, a combined analysis of
$B_s^0\to D_s^{(\ast)+}K^-$ and $B_d^0\to D^{(\ast)+}\pi^-$ allows us to resolve
these problems with the help of the $U$-spin symmetry. In this strategy, where 
$X_d$ has not to be fixed and $X_s$ enters only through a $1+X_s^2$ correction, 
which can be determined through untagged $B_s$ rates, an 
unambiguous value of $\gamma$ can be extracted. The first studies for LHCb 
are very promising \cite{WG5-rep}, and are currently further refined.

\boldmath
\subsection{$B_s\to D\eta^{(')}$, $B_s\to D\phi$, ...\ and $B_d\to D K_{\rm S(L)}$}
\unboldmath
These transitions are the colour-suppressed counterparts of the 
$B_s\to D_s^\pm K^\mp$ channels. If we consider the  CP eigenstates 
$D_\pm$ of the neutral $D$ mesons, we may introduce the following untagged rate 
asymmetry \cite{RF-BCP}:
\begin{equation}
\Gamma_{+-}^{f_s}\equiv
\frac{\langle\Gamma(B_q\to D_+ f_s)\rangle-\langle
\Gamma(B_q\to D_- f_s)\rangle}{\langle\Gamma(B_q\to D_+ f_s)\rangle
+\langle\Gamma(B_q\to D_- f_s)\rangle},
\end{equation}
which allows us to constrain $\gamma$ through 
$|\cos\gamma|\geq|\Gamma_{+-}^{f_s}|$. 
Additional observables are provided by the 
coefficients $C_\pm^{f_s}$ and $S_\pm^{f_s}$ of the  $\cos(\Delta M_qt)$ and 
$\sin(\Delta M_qt)$ terms of the CP asymmetries, respectively. It is convenient
to define the following combinations:
\begin{equation}\label{Cpm-Spm}
\langle C_{f_s}\rangle_\pm\equiv\frac{1}{2}\left[C_+^{f_s}\pm C_-^{f_s}\right],\,
\langle S_{f_s}\rangle_\pm\equiv\frac{1}{2}\left[S_+^{f_s}\pm S_-^{f_s}\right].
\end{equation}
An unambiguous determination of $\gamma$ is then possible with the help of
the following expression \cite{RF-BCP}:
\begin{equation}
\tan\gamma\cos\phi_q=
\left[\frac{\eta_{f_s} \langle S_{f_s}
\rangle_+}{\Gamma_{+-}^{f_s}}\right]
+
\Bigl[\eta_{f_s}\langle S_{f_s}\rangle_--
\sin\phi_q\Bigr],
\end{equation}
where $\eta_{f_s}\equiv(-1)^L\eta_{\rm CP}^{f_s}$ depends on
the angular momentum $L$ of the $D_\pm f_s$ final state.

\boldmath
\subsection{$B_s\to D_\pm K_{\rm S(L)}$ and
$B_d\to D_\pm\pi^0, D_\pm\rho^0, ...$}
\unboldmath
Let us finally have a look at the $b\to d$ counterparts of the  
$B_s\to D\eta^{(')}$, $B_s\to D\phi$, ...\
and $B_d\to D K_{\rm S(L)}$ modes. The relevant interference 
effects are governed by hadronic parameters
$x_{f_d}e^{i\delta_{f_d}}\propto \lambda^2 R_b\approx 0.02$.
Consequently, these decays are not promising for the extraction of $\gamma$. 
However, since their observable combinations $\langle S_{f_d}\rangle_-$, which are 
defined in analogy to (\ref{Cpm-Spm}), satisfy the simple relation 
\begin{equation}
\eta_{f_d}\langle S_{f_d}\rangle_-=
\sin\phi_q + {\cal O}(x_{f_d}^2)
=\sin\phi_q + {\cal O}(4\times 10^{-4}),
\end{equation}
they offer extremely clean extractions of $\sin\phi_q$ \cite{RF-BCP}. In comparison 
with the $B_s\to J/\psi \phi$ and $B_d\to J/\psi K_{\rm S}$ determinations, 
the theoretical accuracy is one order of magnitude higher. Moreover, as no penguin contributions are present, these determinations are very robust with respect to NP 
effects at the decay amplitude level, which is an interesting topic in view of the low 
value of $(\sin2\beta)_{\psi K_{\rm S}}$ and  the ``tension" in the fits of the unitarity 
triangle. In particular, NP effects entering through $B^0_d$--$\bar B^0_d$ mixing or
through tiny effects at the $B\to J/\psi K$ amplitude level (for instance, 
through penguin-like topologies) could be distinguished this way. 

\vspace*{0.5truecm}

\section{Concluding Remarks}
Decays of $B_s$ mesons into final states with charm offer various determinations
of $\gamma$. In these strategies, the $U$-spin flavour symmetry of strong interactions 
provides a powerful tool, allowing us to fully exploit the physics potential of the $B_s$
mesons through a simultaneous analysis of the $U$-spin related $B_d$ decays. 

We also encountered interesting ``by-products" for the $B$-physics programme 
at the LHC, including ways to control the penguin uncertainties in the extractions of 
$\phi_d$ and $\phi_s$ from $B^0_d\to J/\psi K_{\rm S}$ and $B^0_s\to J/\psi \phi$, 
respectively, which seem to be particularly relevant for LHCb upgrade plans, and an 
alternative determination of $\phi_s$ with the help of $B^0_s\to D_s^+D_s^-$. 

The cleanest determinations of the mixing phases $\phi_s$ and $\phi_d$ are offered 
by the pure tree decays $B_s\to D_\pm K_{\rm S(L)}$ and 
$B_d\to D_\pm\pi^0, D_\pm\rho^0, ...$, respectively. These channels would be very 
interesting for the search of NP, but are unfortunately extremely challenging for 
LHCb. On the other hand, they may be accessible at an $e^+e^-$ super-$B$ factory. Detailed experimental feasibility studies in this direction are strongly encouraged.

\end{document}